# Measuring and Predicting Visual Fidelity

Benjamin Watson
watson@northwestern.edu
Dept. Computer Science
Northwestern University
1890 Maple Ave
Evanston, IL 60657 USA
+1 847 491 3710

Alinda Friedman
alinda@ualberta.ca
Dept. Psychology
University of Alberta
Edmonton, Alberta
Canada T6G 2E9
+1 780 492 2909

Aaron McGaffey
mcgaffey@gpu.srv.ualberta.ca
Dept. Psychology
University of Alberta
Edmonton, Alberta
Canada T6G 2E9
+1 780 492 2909

**ABSTRACT**
This paper is a study of techniques for measuring and predicting visual fidelity. As visual stimuli we use polygonal models, and vary their fidelity with two different model simplification algorithms. We also group the stimuli into two object types: animals and man made artifacts. We examine three different experimental techniques for measuring these fidelity changes: naming times, ratings, and preferences. All the measures were sensitive to the type of simplification and level of simplification. However, the measures differed from one another in their response to object type. We also examine several automatic techniques for predicting these experimental measures, including techniques based on images and on the models themselves. Automatic measures of fidelity were successful at predicting experimental ratings, less successful at predicting preferences, and largely failures at predicting naming times. We conclude with suggestions for use and improvement of the experimental and automatic measures of visual fidelity.

**CR Categories:** I.3.7 Three-Dimensional Graphics and Realism, I.3.5 Computational Geometry and Object Modeling

**Keywords:** visual fidelity, model simplification, image quality, naming time, human vision, perception

## 1 INTRODUCTION

Polygonal models, images and the techniques for rendering them are growing steadily in complexity, and with this growth comes a need for visual quality control. For interactive computer graphics applications, fidelity of displayed scenes must be adjusted in real time [Lueb97, Lind96, Redd98]. In many other less interactive applications, models must be simplified to contain fewer polygons, while preserving visual appearance [Garl97, Garl99, Hink93, Ross93, Turk92]. Image generators must determine where and if to add additional image detail [Boli98, Rama99]. Finally, image compression algorithms must preserve appearance while reducing image size [Cosm93, Gers92].

How can visual quality and fidelity be measured? This paper focuses on this question. Ultimately, visual quality can only be assessed by human observers. We compare and contrast three different experimental measures of visual quality: naming times [Wats00], ratings [Cosm93, Mart93] and forced choice preferences. However, the interactive demands of many applications requiring control of visual fidelity do not allow experimentation, which has led many researchers to develop automatic measures of visual fidelity [Boli98, Cign98, Daly93, Lubi93, Rama99]. These measures have then been incorporated into image generation and simplification algorithms [Lind00, Vole00]. We evaluate some of these automatic measures by comparing their results to those of the experimental measures studied herein.

In the following sections, we review the rating, preference, and naming time experimental fidelity measures; present a brief survey of existing automatic fidelity measures; and discuss the small body of computer graphics literature that uses experimental fidelity measures or evaluates automatic fidelity measures. We then present our comparisons and evaluations of several experimental and automatic fidelity measures in the context of model simplification.

## 2 EXPERIMENTAL FIDELITY MEASURES

Ratings and preferences have been widely used in the experimental sciences to obtain relative judgments from human participants. With ratings, observers assign to a stimulus a number with a range and meaning determined by the experimenter. With preferences, observers simply choose the stimulus with more of the experimenter identified quality. Both represent conscious decisions, and so both have proven useful in a wide array of settings, including discomfort ratings in psychiatry, political and popular polling, and the social sciences. With regard to visual fidelity, the experimentally defined meaning or quality of the underlying scales used usually references "quality" or "similarity".

Naming time, the time from the appearance of an object until an observer names it, has a long history of use in cognitive psychology. Existing research has already shown that naming time indexes a number of factors that affect object identification, including the frequency of an object's name in print, the proportion of people who call the object by a particular name and the number of different names in use for it [Vitk95]. Factors of interest to computer graphics researchers include viewpoint [Joli85, Palm81], familiarity [Joli89] and structural similarity [Bart76, Hump95]. In work of particular interest for this study, researchers have shown repeatedly that natural objects take longer to name than manmade artifacts [Hump88]. They hypothesize that natural objects are structurally more similar to one another, requiring more disambiguation than artifacts.

## 3 AUTOMATIC FIDELITY MEASURES

Although these experimental measures of visual fidelity can be quite effective, time or resources often do not allow their use. In such cases researchers and application builders often turn to automatic measures of visual fidelity.

For level of detail control, researchers estimate error by tracking the deviation of geometry in the image plane [Lueb97, Lind96], and possibly modulating the importance of this error with knowledge of human perception [Redd98]. On the other hand, model simplification researchers have long used three dimensional (3D) measures of distance [Ross93], curvature [Hink93, Turk92], or volume [Lind99] since one typically does not know what part of the model an user may be observing, and these measures are view independent. Lindstrom [Lind00] has measured fidelity for simplification by taking virtual snapshots of the model being simplified from several view points, and then measuring the difference between the snapshots taken before and after the simplification with mean squared error (MSE) (see below). Although it was not used in actual simplification, Cignoni, Rocchini, and Scopigno [Cign98] have offered the Metro tool, which allows users to evaluate the quality of already simplified models with 3D measures of distance and volume.

In the fields of image generation and compression, researchers have focused on view dependent automatic fidelity measures that compare the quality of images. The MSE measure simply finds the mean of the squared pixel by pixel differences between the original and approximate images (often the differences are normalized by the squared value of the pixel in the original image). However, recently several shortcomings of MSE were noted [Giro93] and more complex measures were built based on numerical models of the early stages of the human visual system [Boli98, Daly93, Lubi93]. These were then used to evaluate image compression algorithms and incorporated into image generation algorithms [Boli98, Rama99, Vole00].

## 4 PREVIOUS FIDELITY EXPERIMENTS

The study of visual fidelity measures and their is are just beginning to make their way into computer graphics research. Rushmeier, Rogowitz and Piatko [Rush00] used a fine grained, one dimensional experimental rating (or scaling) system to evaluate the effects on fidelity of approximations in geometry and texture. They found indications that the ability of texture to hide approximations in geometry depends on the coarseness of the original geometry. Pellachini, Ferwerda and Greenberg [Pell00] used similarity ratings combined with multidimensional scaling and magnitude estimation to derive a perceptually equidistant gloss space.

An initial perceptual evaluation of the automatic fidelity measure designed by Daly was performed by Martens and Myszkowski [Mart93]. They found a high correlation between the Daly measures and observer ratings of texture masked objects. Previously we [Wats00] used naming times as an experimental fidelity measure to examine the effects of model simplification. After duplicating the natural/manmade effect discussed above and confirming that naming times were sensitive to simplification, we turned to an evaluation of several automatic fidelity measures. We found that at severe simplifications, the automatic measure designed by Bolin (BM) [Boli98] was the most reliable, with MSE and maximum 3D distance also fairly reliable. However, at more moderate simplifications none of the automatic measures reliably modeled naming time.

## 5 EXPERIMENTAL MEASURE STUDY

Our evaluation of the naming time, rating, and preference fidelity measures took the form of an experiment using these measures as the dependent outcomes. This experiment had two goals: to learn about the relative strengths and weaknesses of these measures in their responses to model and image fidelity, and to provide an experimental test bed for our evaluation of automatic fidelity measures in the following section.

### 5.1 Methods

Here we outline experimental methodology and detail our experimental stimuli. For full detail, please see the appendix.

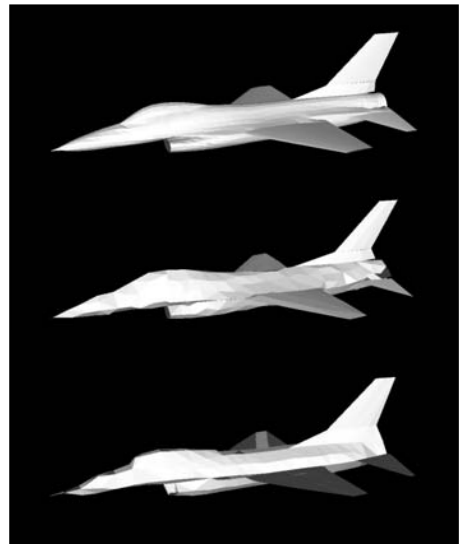

**Figure 1:** One stimuli from the experimental set. At the top is the original, the middle Qslim 80%, at the bottom Vclust 80%.

Stimuli were created from 36 3D polygonal models (31 in the public domain; 5 from a commercial source). None contained color, texture, material, or vertex normal information. Half the models represented manmade artifacts and the other half were representations of animals. Each of these models was simplified using two simplification algorithms (Vclust [Ross93] and Qslim [Garl97]) resulting in two levels of simplification each. We chose these algorithms because they are widely used and according to prevailing opinion, produce models differing widely in visual fidelity. Thus this experiment had three independent variables: *simplification type* (Vclust vs. Qslim), *simplification level* (three levels including unsimplified), and *object type* (animals vs. artifacts). These were varied within participants.

Models were simplified in two stages. First, Qslim was used to simplify all models to the number of polygons contained in the smallest model in the set (3700 ±50). We refer to these as the "standards" (0% simplification), and label a member of this set *s*. Second, the standards were simplified using Qslim and Vclust by removing 50% and 80% of the original 3700 polygons. We refer to members of the resulting four model sets as *q5*, *q8*, *v5* and *v8*. There were thus five examples of each of the 36 objects, for a total of 180 stimuli.

Each stimulus image was uniformly scaled to 591 pixels in width and displayed in the center of the screen. The rating and preference task stimuli each consisted of two exemplars of a single object model that were scaled to 400 pixels in width and displayed side-by-side, each centered within a 512(w) x 768(h) pixel space. Figure 1 shows a stimulus simplified at 80% by Qslim and Vclust.

*Naming task*. Participants were asked to name each object as quickly and accurately as they could. They were told that some pictures would be simplified representations and were shown printed examples.

*Rating task*. All four simplified exemplars of an object were rated against the standard ((*s*, *q5*), (*s*, *q8*), (*s*, *v5*) and (*s*, *v8*)). Each participant rated all 36 objects once at each simplification type and level. Stimuli were presented in a random order.

Participants were told that on each trial their task was to rate the likeness of the picture on the right against the standard picture on the left, using a 7-point scale. They had four practice trials.

*Preference task*. Exemplars of both simplification types were compared at the same simplification level (e.g. (*q5*, *v5*) or (*q8*,

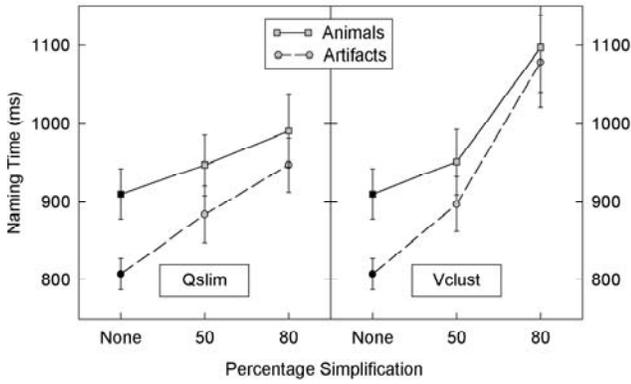

**Figure 2:** Naming times as a function of simplification type, simplification level, and object type.

| Variable | Avg By | ANOVA |
|---|---|---|
| object type | participants | $F(1,35) = 10.24$ |
| simp level | participants | $F(1,35) = 13.59$ |
| simp level | objects | $F(1,33) = 13.80$ |

**Table 1:** 2 way analysis on naming times averaged over simp type. All effects $p<.05$.

| Variable | Avg By | ANOVA |
|---|---|---|
| simp type | participants | $F(1,35) = 5.29$ |
| simp level | participants | $F(1,35) = 13.59$ |
| simp level | objects | $F(1,33) = 13.80$ |
| stype x slevel | participants | $F(1,35) = 4.70$ |

**Table 2:** 3 way analysis on naming times without standard models. All effects $p<.05$.

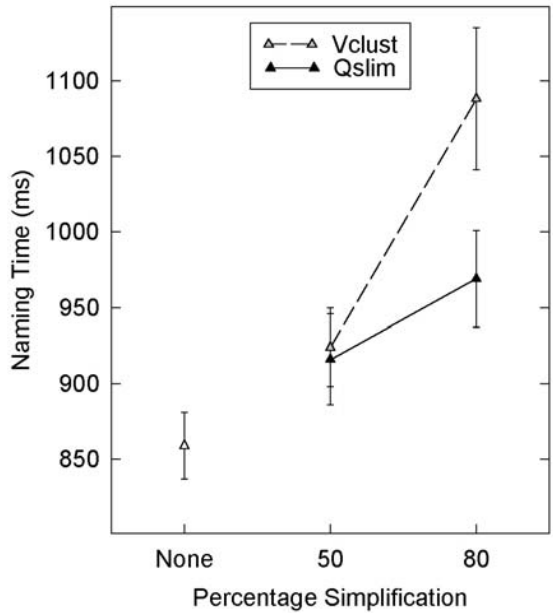

**Figure 3:** Naming times averaged across object type as a function of simplification type and level.

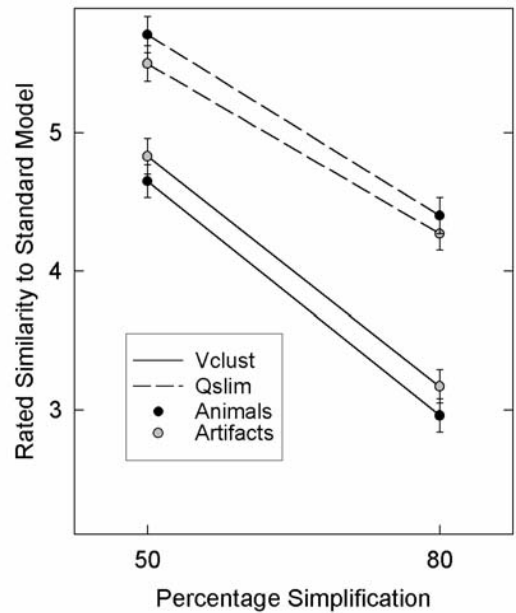

**Figure 4:** Ratings by simplification type, simplification level, and object type.

*v8*)). There were 36 objects, each with two simplification types and two simplification levels, for a total of 72 comparisons. The left-right position of the Qslim (Vclust) example was distributed evenly throughout the trials. Participants had four practice trials. Participants were asked to choose which picture in the set was a better example of each object.

### 5.2 Results

#### 5.2.1 Naming Times

Figure 2 shows the mean naming times as a function of object type, simplification algorithm, and simplification level. It can be seen that all three factors affected performance: animals were named more slowly than artifacts, naming times were longer with increasing simplification, and naming times were longer with Vclust (see Table 1). There were no interactions between object type and simp level. Reassuringly, this replicates the main trends of our earlier study [Wats00]. In the only interaction, the effect of simp type varied with simp level (see Table 2). Figure 3 shows the data averaged over type of model. Clearly the Vclust algorithm was much more devastating to naming times than Qslim at the higher levels of simplification.

We corroborated these observations with analyses of variance (ANOVAs) on the naming time means averaged two ways. For details on these analyses please see the appendix.

#### 5.2.2 Ratings and Preferences

Rating results are shown in Table 3, averaged two ways. Figure 4 shows the average similarity ratings as a function of object type, simplification type, and simplification level. Participants were sensitive to simplification level and rated the 50% simplified objects closer to the "ideal" than the 80% simplified objects (5.2 versus 3.7). Second, they also clearly thought that the Qslim-simplified objects were closer to the ideal than were the Vclust-simplified objects (5.0 versus 3.9). Third, simplification type interacted with simplification level; similar to the naming time data, there was less of a difference between the algorithms when objects had been simplified to 50% (5.6 versus 4.7 for Qslim and Vclust, respectively) than when they had been simplified to 80% (4.3 versus 3.1).

| Variable | Avg By | ANOVA |
|---|---|---|
| simp type | participants | F(1,35) = 243.56 |
| simp type | objects | F(1,33) = 100.97 |
| simp level | participants | F(1,35) = 264.29 |
| simp level | objects | F(1,33) = 388.86 |
| stype x slevel | participants | F(1,35) = 32.23 |
| stype x slevel | objects | F(1,33) = 11.75 |
| stype x otype | participants | F(1,35) = 29.51 |

**Table 3:** 3 way statistical analysis on ratings. All effects p<.05.

| Variable | Avg By | ANOVA |
|---|---|---|
| object type | participants | F(1,35) = 79.68 |
| object type | objects | F(1,33) = 5.25 |
| simp level | objects | F(1,33) = 18.20 |

**Table 4:** 2 way analysis on preferences. All effects p<.05.

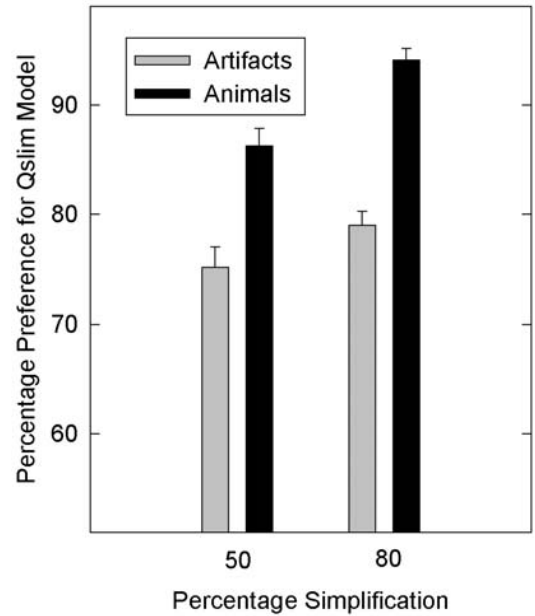

**Figure 5:** Preferences for Qslim by simp level and object type.

In all of these respects, ratings results were similar to naming time results. However, ratings and naming times differed in their response to object type. Ratings did not respond simply to object type, and in fact there was an interaction between object type and simp type: the animal models were rated closer to the standard when they had been simplified using the Qslim algorithm (5.1 versus 4.9 for animals versus artifacts, respectively), but the artifacts were rated as being closer to the standard when they had been simplified using Vclust (3.8 versus 4.0 for animals versus objects).

In the preference results, there were main effects for both object type and simplification level (see Table 4 and Figure 5). Essentially, the preference for Qslim-simplified stimuli was greater for the animal models than for the artifact models (90.1% versus 77.0%), and it was greater for 80% objects than for the 50% objects (86.5% versus 80.6%).

## 6 AUTOMATIC MEASURE STUDY

We now turn our attention to automatic measures of visual fidelity, and their ability to predict experimental measures of fidelity provided by human observers. Such automatic measures, if effective, could be quite useful in evaluating the effectiveness of various algorithms -- and if efficient enough, might even be incorporated into the algorithms themselves. We examine three tools for measuring fidelity: an implementation of the image comparison algorithm described by Bolin and Meyers [Boli98] (BM), mean squared image error (MSE), and the Metro tool from Cignoni, Rocchini and Scopigno [Cign98].

### 6.1 Methods

Both BM and MSE accept as input an ideal image and an approximate image, and return summary measures of the difference between these images. MSE returns a single number as its estimate. BM returns a difference image, with the value at each image location estimating the ability of viewers to perceive the local difference between the images. Since we require a single value summarizing image fidelity, we use the average of all the local values contained in the difference image. For both MSE and BM, the images used were the same images used in experimentation.

Metro accepts as input two similar 3D polygonal models, and as a result is not sensitive to viewpoint. It returns rough estimates of the difference in volume between the two models. It also samples the two surfaces at multiple points, and measures the distance from each point on the first (pivot) model to the surface of the other model. It returns three summaries of these distance measures. The first is the mean of these distances, obtained by normalizing the sum of the distances with the surface area of the pivot model. The second simply squares each of the summed distances before normalization. The third is the maximum of the measured distances. Metro returns its summaries in model coordinates, as well as in coordinates normalized by the diagonal of the pivot model bounding box and the diameter of the smallest sphere that encloses the pivot model.

Our evaluation of Metro's fidelity measures includes the volume difference (MetroVol), as well as each of the mean (MetroMn), mean squared (MetroMSE) and maximum (MetroMax) summaries. All three distance summaries were normalized by the diagonal of the pivot model bounding box. For the maximum summary, we used a Metro option that returned the Hausdorf distance, that is, that found the maximum of two-sided distance measurements both from the first model to the second, and the second to the first.

We found four sets of automatic fidelity measures for each of the 36 models in the experimental set. If, for a given model, $s$ is the standard, $q5$ and $q8$ are versions of $s$ simplified by Qslim 50% and 80% respectively, and $v5$ and $v8$ are versions of $s$ simplified by vertex clustering 50% and 80% respectively, then we found sets of fidelity measures for each of the following four model pairs: $(s,q5)$, $(s,q8)$, $(s,v5)$ and $(s,v8)$. For each of these pairs, the set of fidelity measures included BM, MSE, MetroMn, MetroMSE, MetroMax and MetroVol. For Metro, we always used $s$ as the pivot.

To each fidelity measure in each model pair set we compared experimental measures. For naming times, we used the time for the non-standard model in the pair (e.g. for $(s,q5)$, we used the time it took to name $q5$, or $name(q5)$). For ratings, we used the rating of the non-standard model in comparison to the standard (e.g. $rate(s,q5)$).

Automatic measures compared to experimental preference measures took a special form. Typically persons will compare two stimuli for quality by judging which of the two is closer to a visually presented or completely cognitive ideal. Therefore the automatic measures we compared to experimental preferences were constructed from the previous measured pairings, and took the form $p5 = (meas(s,q5) - meas(s,v5))$ and $p8 = (meas(s,q8) - meas(s,v8))$, where *meas* is one of the six measures we evaluated. *p5* and *p8* predict preference among the 50% and 80% simplified models, respectively, with a positive result predicting a preference for Vclust, a negative result for Qslim. We also compared naming times and ratings to *p5* and *p8*. These comparisons used the differences in naming times and ratings across simplification type (e.g. $(name(q5) – name(v5))$ and $(rate(s,q5) – rate(s,v5))$).

| Automatic Measure | Naming Times | | | | | | Ratings | | | | | |
|---|---|---|---|---|---|---|---|---|---|---|---|---|
| | All Models | | Animals | | Artifacts | | All Models | | Animals | | Artifacts | |
| | Qslim | Vclust | Qslim | Vclust | Qslim | Vclust | Qslim | Vclust | Qslim | Vclust | Qslim | Vclust |
| BM | -0.07 | ***0.30*** | -0.07 | 0.21 | -0.03 | ***0.41*** | *-0.62* | *-0.60* | *-0.43* | *-0.54* | *-0.72* | *-0.67* |
| MSE | 0.07 | ***0.31*** | 0.02 | 0.14 | 0.18 | ***0.48*** | *-0.67* | *-0.71* | *-0.68* | *-0.71* | *-0.74* | *-0.77* |
| MetroMn | 0.03 | ***0.31*** | 0.00 | 0.24 | 0.10 | ***0.38*** | *-0.65* | *-0.77* | *-0.77* | *-0.78* | *-0.66* | *-0.77* |
| MetroMSE | -0.04 | ***0.25*** | -0.20 | 0.27 | 0.06 | 0.22 | *-0.46* | *-0.55* | -0.21 | *-0.53* | *-0.56* | *-0.60* |
| MetroMax | -0.05 | ***0.27*** | -0.16 | 0.26 | 0.04 | 0.28 | *-0.60* | *-0.73* | *-0.52* | *-0.75* | *-0.66* | *-0.72* |
| MetroVol | 0.19 | 0.14 | -0.07 | 0.08 | ***0.41*** | 0.19 | -0.21 | -0.13 | *-0.58* | *-0.34* | 0.00 | -0.04 |

**Table 5:** Correlations of naming times and ratings to automatic fidelity measures.

| Fidelity Measures | Simp Type | Simp Level | SType x SLevel | SType x OType | Three Way |
|---|---|---|---|---|---|
| Naming | *5.29* | 13.80 | *4.70* | | |
| Rating | 100.97 | 388.86 | 11.75 | *29.51* | |
| BM | 11.73 | 108.08 | 6.31 | | |
| MSE | 78.31 | 100.12 | 37.55 | | |
| MetroMn | 56.48 | 192.71 | 32.27 | 8.02 | 8.18 |
| MetroMSE | 23.58 | 135.08 | 14.72 | 8.67 | 7.03 |
| MetroMax | | 32.86 | | | |
| MetroVol | | 6.68 | 7.82 | | |

**Table 6:** Significant ANOVAs for naming times, ratings and automatic fidelity measures. Italics represent participant analyses.

## 6.2 Results

Table 5 shows automatic fidelity measure correlations to naming times and ratings used to judge quality with (at least implicit) reference to an ideal. Each correlation measure reflects comparisons for both simplification levels within a simplification type. Where correlations are presented in bold, the associated automatic measure accounts for a marginally significant ($p < 0.1$) proportion of the variation in the experimental measure. Where they are also italicized, the automatic measure accounts for a significant ($p < 0.5$) proportion of experimental variation.

All automatic measures with the exception of MetroVol were very successful predictors of quality as judged by ratings. Correlations were quite high, with ANOVAs indicating that a statistically significant portion of experimental variance was accounted for. Note that correlations are consistently negative, since low automatically measured error correlates consistently with high experimental ratings. Correlations are slightly worse for animals as opposed to artifacts, and for Qslim as opposed to Vclust.

The automatic measures were much less successful at predicting quality as judged by naming times. Correlations were in this case generally positive, since low automatically measured error correlates to short naming times. The most successful automatic fidelity measures were BM, MSE and MetroMn. The striking failures here are the consistently low correlations for Qslim, and to a lesser degree, the lower correlations for animals, echoing the same trends in the ratings correlations.

We performed in-depth analyses of the automatic measures by treating their results as dependent variables in ANOVAs much like those used for the experimental measures, with simplification type, simplification level, and object type as independent variables. We present these results in Table 6. Table values in italics represent F values from analyses averaged across objects for each participant, rather than averaged across participant for each object. We graph the means for two of the better measures, BM and MetroMn, by objects in Figures 6 and 7, and show for comparison naming time and ratings graphs averaged over participants for each object. All measures, whether automatic or experimental, were significantly affected by simplification level. Most measures were significantly affected by simplification type and the interaction of simplification type and level. The effect of object type, however, differed greatly across the measures, whether experimental or automatic.

Table 7 shows automatic fidelity measure correlations to preferences and naming time and rating differences used to judge which of two stimuli has superior quality. Each correlation measure again reflects comparisons for both simplification levels. Where correlations are presented in bold, the automatic measure accounts for a marginally significant ($p < 0.1$) proportion of the variation in the experimental measure, where they are also italicized, the automatic measure accounts for a significant ($p < 0.5$) proportion of experimental variation. The automatic measure differences are negative if Qslim has less error, while rating differences and preferences are positive if Qslim is rated more highly or preferred, giving negative correlations. Since naming time differences are negative if Qslim produces the more recognizable model, correlations to it are largely positive. In general, the automatic measures correlated quite well to experimental preferences, less well to differences in ratings, and quite poorly to differences in naming times. Again correlations were worse for animals than for artifacts.

## 7 DISCUSSION

In this section we review our experimental and automated findings, make some recommendations on the use of fidelity measures, and provide some suggestions as to how automatic fidelity measures and the applications that use them might be improved.

## 7.1 Limitations

However, before we do so, we should note the limitations of our studies. We begin with a consideration of our stimuli. First, we have limited ourselves to the study of one almost optimal view of each object. Second, this study has focused on approximations made in model geometry, rather than in the illumination model, model texture, or in attributes such as color or per-vertex normal vectors. In addition, we have focused on recognition of objects presented in isolation, rather than in more natural scenes

**Table 7:** Correlations of preferences, naming time differences, and rating differences to automatic fidelity measures.

| Automatic Measure | Naming Diffs | | | Rating Diffs | | | Preferences | | |
|---|---|---|---|---|---|---|---|---|---|
| | All | Anims | Artifs | All | Anims | Artifs | All | Anims | Artifs |
| BM | 0.21 | 0.23 | 0.23 | ***-0.36*** | -0.23 | ***-0.38*** | ***-0.37*** | -0.27 | ***-0.35*** |
| MSE | ***0.26*** | 0.15 | ***0.37*** | ***-0.44*** | -0.25 | ***-0.54*** | ***-0.33*** | ***-0.42*** | -0.27 |
| MetroMn | 0.18 | 0.20 | 0.21 | ***-0.42*** | -0.21 | ***-0.47*** | ***-0.42*** | ***-0.41*** | -0.32 |
| MetroMSE | 0.04 | 0.17 | -0.03 | -0.21 | -0.25 | -0.15 | ***-0.27*** | ***-0.42*** | -0.16 |
| MetroMax | 0.13 | 0.19 | 0.14 | ***-0.41*** | -0.16 | ***-0.45*** | ***-0.43*** | ***-0.40*** | ***-0.34*** |
| MetroVol | -0.06 | 0.17 | -0.17 | -0.05 | 0.19 | -0.15 | -0.04 | 0.16 | -0.11 |

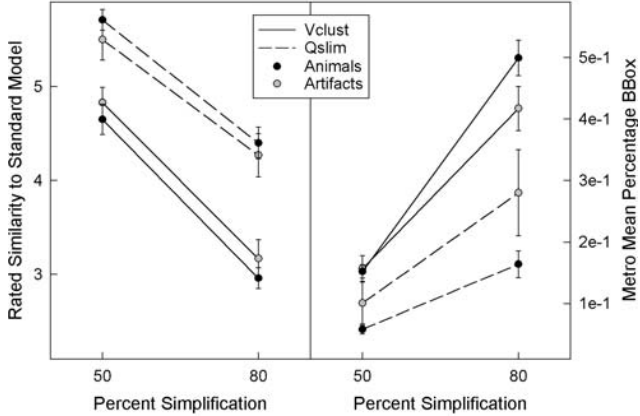

**Figure 6:** MetroMn response to simplification type, simplification level, and object type, vs. ratings by object.

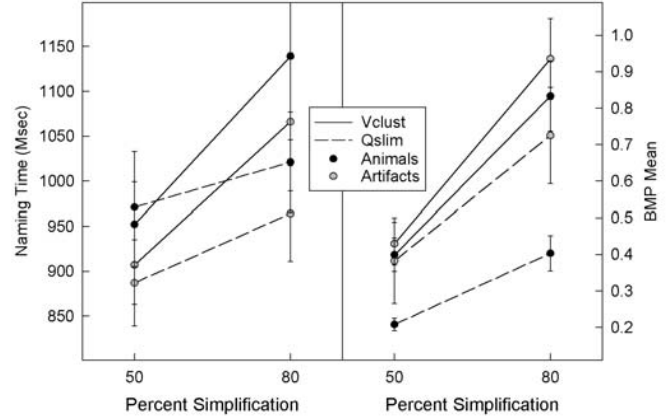

**Figure7:** BM response to simplification type, simplification level, and object type, vs. naming time by object.

containing several objects in their context. Finally, models are often used in interactive applications with viewpoint and model motion, while all of the stimuli presented and studied here were static. Removing these limitations in further studies would certainly increase the generality of our results. At the same time, such changes would increase variation in those results, and stiffen the challenge posed to the various automatic measures of visual fidelity, which would have to model a more complex experimental response. For example, the introduction of viewpoint and non-geometric experimental factors would certainly reduce the effectiveness of the Metro measures, at least in their current form.

In order to limit the scope of our experimentation, we also made choices in the use of our automatic measures. In particular, we chose BM as a quickly executing representative of those measures that model the early stages of human vision. But BM is a specialization of other slower measures ([Daly93, Lubi93]) that might be more effective (though BM was proved very effective in these results). BM and related measures were also developed for stimuli more complex (and more challenging) than those used here. Difference image summarizations other than the averaging used here might also increase measure effectiveness.

### 7.2 Confirmations

As we have noted above, our naming time results were largely in agreement with the results we obtained earlier in [Wats00]. We also found that simp level has the effect one would intuitively expect on the rating and preference measures. In agreement with prevailing opinion, Qslim was by all measures a more effective simplification tool than Vclust. Many have conjectured that simplification techniques show their mettle at low polygon counts. These results are in agreement with that hypothesis, with a simp level and simp type interaction showing that there is little difference between Qslim and Vclust at 50% simplification, a large difference at 80%.

### 7.3 Surprises

Ratings and preferences indicated that though Qslim is generally a better simplifier than Vclust, it simplifies animals most effectively. This may indicate that a specialization of Qslim for more regularly curved (or planar) surfaces is possible. On the other hand, Vclust is more effective when simplifying artifacts – a hint that Vclust's regular sampling approach is most effective when used with models typical of CAD/CAM applications, which contain many coplanar polygons and regularly curved surfaces.

Naming times did not respond to object type with the same complexity as ratings and preferences, instead, they were uniformly longer for animals. This is most likely a clue to the different natures of these experimental measures: while naming times probe subconscious perception from the low through the higher cognitive levels, ratings and preferences seem to sample very low level processes, avoiding the natural/manmade effect. (However, ratings and preferences are notoriously vulnerable to higher level, conscious cognitive qualities assigned to the axis of comparison).

In line with these differences in the experimental results, the automatic measures were poor predictors of naming times, but excellent predictors of experimental ratings, preferences, and to a lesser extent, differences in ratings. BM, MSE, and MetroMn were particular success stories in this respect. Obviously the differing experimental responses to object type played a role in these correlative trends.

However, correlations were low even within simplification and object type, where only simplification level was varying. We see two possible reasons for this. First and most simply, naming times are very variable, much more so than ratings and preferences. Obtaining good correlations to them within simplification and object type may require the use in experiments of larger model stimuli libraries and more participants. Second and more provocative, although it is certain that in general, increasing simplification increases naming times; we noticed that for several models, increasing simplification *reduced* naming times. We will call this the *distillation effect*. There are precedents for this in the psychology literature [Ryan56, Edel99]. The basic notion is that by removing detail that allows fine grained identification of an object, the speed of coarse grained, categorizing identification is improved. The distillation effect seems to occur particularly often for animal models simplified with Qslim, and may explain some of the negative correlations (again, within simplification and object type) in Table 5. Automatic measures do not model this effect, reducing correlations.

## 7.4 Implications

*For simplification.* Our results indicate that simplification effectiveness varies by all experimental measures as a function of object type. This suggests the possibility of simplification algorithms that specialize in, or adapt to, different types of models. As simplification researchers continue their work, they should pay particular attention to the quality of their models at low polygon counts. Our results also suggest that mean distance is a more important heuristic for simplification than maximum distance. The rating and preference measures are well modeled by the automatic measures reviewed here, which should prove useful when comparing algorithms, or even during the process of simplification itself.

*For use of experimental measures.* All of the measures vary in the degree of explicit visual comparison they require of the viewer. With preferences, this comparison is very explicit. With ratings, the comparison is to some (at least cognitively imagined) visual ideal. Comparison may be involved in the process probed by naming times, but it is certainly not an explicit comparison between two visual images. It may be telling that this least comparative of experimental measures was also most poorly modeled by the automatic measures.

What is fidelity? Is it visual similarity to the original? Or is it successful communication of the original concept? The notion of fidelity most relevant in the current application should indicate the experimental measure that is most appropriate. During the processes of simplification, image compression and generation, the goal is typically one of appearance preservation in the face of each of a long series of minor alterations. Preferences, ratings, and their correlating automatic measures are probably the most appropriate indices for these applications. However, when making cross algorithm comparisons, the compared images or models are the result of very different, very long series of these alterations, and the appropriate notion of fidelity is less clear.

It is intriguing to note that in almost all computer graphics applications, users *never* make an explicit visual comparison. At most, they compare a currently displayed example with a previously displayed one. In highly interactive applications, this comparison, if it indeed occurs, is certainly cursory at best. In these sorts of settings, the naming time measure might be most appropriate, and the distillation effect, if it indeed exists, most effectively exploited. It is also intriguing to imagine a non-photorealistic pursuit of the distillation effect in its extreme.

*For automatic measures.* Many of these measures can be used for purely numerical ends, ensuring for example that a given approximation does not deviate from the original by more than some constant error. We do not consider such applications here.

Our results indicate that MetroVol is a poor predictor of visual fidelity and quality as indexed by any of our experimental measures, at least at the levels of geometric simplification (3700 polygons and below) examined here. BM, MSE, and MetroMn were excellent predictors of fidelity as measured by ratings and to a lesser extent preferences and rating differences. Unfortunately, we found no fully reliable predictors of the conceptual sort of fidelity measured by naming times. For now, the best automatic predictor of naming times and their differences is MSE, with BM and MetroMn coming very close behind. Given the poor correlations of all three of these measures with Qslim, these naming time predictors must be used with extreme skepticism, if at all.

*For future work.* These results raise many intriguing questions. First, do they generalize? We are currently investigating how well these results hold across differ viewpoints, and would like to examine the effects of both background and interactive motion. The element of comparison embodied both by these measures and typical graphics applications clearly needs further research, as does the hypothesized distillation effect. Obviously our automatic measures must improve their ability to model naming times. This will require understanding and modeling object type effects. In the long run, research into the object type and distillation effects may lead to new simplification algorithms.

## 8 CONCLUSION

This paper described our research into the experimental and automatic measurement of visual fidelity. Measuring visual fidelity is fast becoming crucial in the fields of model simplification, level of detail control, image generation, and image compression. In our study, we manipulated fidelity by applying two different model simplification algorithms to 36 polygonal models, divided into models of animals and manmade artifacts, producing approximating models at two different polygon counts. We examined the visual fidelity of these models with three different experimental measures: naming times, ratings, and preferences. All the measures were sensitive to the type of simplification algorithm used and the amount of simplification, however they responded differently to model type. We then analyzed model visual fidelity with several automatic measures of visual fidelity. These automatic measures proved to be good predictors of ratings and preferences, but only mediocre predictors of naming times.

## 9 ACKNOWLEDGEMENTS

Oscar Meruvia wrote indispensable 3D viewing software. Josh Anon helped with simplifications and quality predictors. We thank Greg Turk for his models and geometry filters. Peter Lindstrom participated with a thought provoking correspondence, and by assisting in finding relevant code. Oleg Veryuvka and Lisa Streit provided useful implementations of simple image quality metrics. Stanford University was the source of the bunny model. This research was supported by NSERC grants to the first two authors.

## 10 APPENDIX: EXPERIMENTAL DETAILS

*Experimental methodology.* Participants performed the naming, rating, and forced-choice preference tasks on the same set of items during a single session. All participants completed the naming task first because seeing a stimulus once reduces its subsequent naming time [Joli85]. Similarly, all participants performed the rating task prior to the preference task because it was possible that performing the preference task first could contaminate rating judgments by increasing the subjective distance between the less preferred object and the standard.

The virtual field of view used in forming stimuli was always 40 degrees and the virtual eye point was always at a distance that was twice the length of the bounding box. Views were generally directed towards the mean of a model's vertices, but 14 models required centering corrections because vertex distributions were not uniform. Each model was interactively rotated so that it was displayed in a canonical 3/4 view that revealed a reasonable level of detail across the models [Palm81]. Each model was illuminated with one white (RGB=[1,1,1]) light located at the eye point. All models were assigned the same white color and flat shaded, and displayed on a black background.

The images were displayed on a 17-inch Microscan CRT, with participants sitting approximately 0.7 m from the display. Participants performed the naming task by speaking into a hand-held microphone. Responses for the rating and preference tasks were entered on the computer keyboard. Thirty-six undergraduate volunteers from the University of Alberta pool participated in the experiment.

For the naming task, stimuli were organized into six groups of six stimuli each. There were three groups for each simplification algorithm, and within algorithms, one group for each level of simplification (0%, 50%, and 80%). Each group contained three animal models and three artifact models. Stimuli were cycled through the groups such that across participants, each stimulus appeared once in each of the six experimental conditions (2 simp type x 3 simp level).

Each participant saw all 36 models only once; 12 were standards, 12 were simplified using Qslim, and 12 were simplified using Vclust; 6 of each of the simplified models were seen at 50% simplification and 6 were seen at 80% simplification. There were eight practice trials. On each trial, the experimenter pressed the space bar, a fixation cross appeared for 750 ms, the picture appeared on the screen, the participant named the picture, and the picture disappeared as soon as a name was said. Naming times were recorded from stimulus onset to the participant's response.

For the rating task, on each trial the participant pressed the space bar, and after a delay of 250 ms the standard and comparison pictures appeared on the screen and disappeared as soon as a rating was entered.

For the preference task, subjects pressed the "A" and "K" keys to choose the left and right stimuli, respectively. The participant pressed the space bar; after a delay of 250 ms the pictures appeared and then disappeared as soon as a preference was entered.

The models: ant, bear, bicycle, blender, bunny, camera, car, chair, cow, dinosaur, dog, dolphin, dump truck, elephant, fighter jet, fish, helicopter, horse, kangaroo, lion, microscope, motorcycle, piano, pig, plane, raven, rhino, sandal, shark, ship, skateboard, snail, sofa, spider, tank, tomgun.

*Analysis of experimental measure results.* Three kinds of trials were excluded from naming time analyses. First, we excluded naming times measured during spoiled trials (e.g., trials in which participants failed to trigger the microphone with their first vocalization – 4.6% of all trials). Second, we excluded naming times from trials in which a participant's response was an error (e.g., calling a picture of a sandal a "rocket" – 0.3% of all trials). Finally, we computed the overall mean of the remaining naming times and excluded trials that were more than 3 standard deviations longer than this average. These outliers comprised only 1.5% of the remaining trials.

For naming times and ratings, examining the relationship of object type to simp level required averaging over simp type for a two way analysis, because unsimplified objects were necessarily the same for both the Qslim and Vclust. Additional three way analyses were performed by excluding the unsimplified objects.

Most analyses used two ANOVAs, one averaged over objects (the participant analysis) and one averaged over participants (the object analysis). For the participant ANOVA on the preference data, we counted the frequency of times that each participant chose the Qslim-simplified model in each of the four object type and simplification level conditions, and converted the results to percentages. For the item analysis, we counted the frequency of participants who chose the Qslim model for each of the objects in each of the conditions.